\documentclass[aps,superscriptaddress,twocolumn,twoside,floatfix,pra,nofootinbib,a4paper]{revtex4-2}

\pdfoutput=1

\usepackage{times}
\usepackage{dsfont}
\usepackage{bm} 
\usepackage{amsfonts}
\usepackage{amsmath}
\usepackage{amssymb}
\usepackage{amsthm}
\usepackage{color}
\usepackage{enumerate}
\usepackage{multirow}
\newcommand{\stkout}[1]{\ifmmode\text{\sout{\ensuremath{#1}}}\else\sout{#1}\fi}
\usepackage{latexsym}
\usepackage{mathrsfs}
\usepackage{verbatim}
\usepackage{float}
\usepackage{pifont}
\usepackage{caption}
\usepackage{empheq}
\usepackage{IEEEtrantools}

\usepackage{microtype}

\usepackage[dvipsnames]{xcolor}

\usepackage{natbib}
\usepackage[colorlinks=true,linkcolor=blue,citecolor=magenta,urlcolor=blue]{hyperref}

\usepackage{cleveref} 

\usepackage[many]{tcolorbox}

\newtcolorbox[auto counter]{framefloat}[2][]{title=Box~\thetcbcounter: #2,,fonttitle=\bfseries, boxsep=0mm,boxrule=1pt,colframe=black,colback=white,coltitle=black,float=t!,#1}

\definecolor{maroon}{cmyk}{0,0.87,0.68,0.32}

\DeclareMathOperator{\Tr}{tr}

\usepackage{physics} 

\newcommand{\trnorm}[1]{\norm{#1}_{1}}
\newcommand{\vect}[1]{\boldsymbol{#1}}

\newcommand{\id}{\mathds{1}}

\newcommand{\etal}{\textit{et al.\ }}

\newcommand{\EA}{\mathrm{EA}}

\usepackage{array,mathtools,amssymb,booktabs}
\newcolumntype{C}{>{$}c<{$}}
\AtBeginDocument{
\heavyrulewidth=.08em
\lightrulewidth=.05em
\cmidrulewidth=.03em
\belowrulesep=.65ex
\belowbottomsep=0pt
\aboverulesep=.4ex
\abovetopsep=0pt
\cmidrulesep=\doublerulesep
\cmidrulekern=.5em
\defaultaddspace=.5em
}

\usepackage{epsfig}
\usepackage{graphicx}
\graphicspath{{figures/}}
\usepackage{subfig}
\usepackage{caption}
\newcommand{\liq}{Laboratoire d'Information Quantique, Universit\'e libre de Bruxelles (ULB), Belgium}

\newcommand{\iqoqi}{Institute for Quantum Optics and Quantum Information -- IQOQI Vienna, Austrian Academy of Sciences, Boltzmanngasse 3, 1090 Vienna, Austria}

\newcommand{\iasp}{Institute for Atomic and Subatomic Physics, Vienna University of Technology, 1020 Vienna, Austria}

\begin{document}

\title{Entanglement in prepare-and-measure scenarios: many questions, a few answers}

\author{Jef Pauwels}
\affiliation{\liq}

\author{Armin Tavakoli}
\affiliation{\iqoqi}	
\affiliation{\iasp}

\author{Erik Woodhead}
\affiliation{\liq}
	
\author{Stefano Pironio}
\affiliation{\liq}

\date{January 28, 2022}

\begin{abstract}
Entanglement and quantum communication are paradigmatic resources in quantum information science leading to correlations between systems that have no classical analogue.  Correlations due to entanglement when communication is absent have for long been studied in Bell scenarios. Correlations due to quantum communication when entanglement is absent have been studied extensively in prepare-and-measure scenarios in the last decade. Here, we set out to understand and investigate correlations in scenarios that involve both entanglement and communication, focusing on entanglement-assisted prepare-and-measure scenarios. In a recent companion paper [\href{https://arxiv.org/abs/2103.10748}{arXiv:2103.10748}], we investigated correlations based on unrestricted entanglement. Here, our focus is on scenarios with restricted entanglement.  We establish several elementary relations between standard classical and quantum  communication and their entanglement-assisted counterparts. In particular, while it was already known that bits or qubits assisted by two-qubit entanglement between the sender and receiver constitute a stronger resource than bare bits or qubits, we show that higher-dimensional entanglement further enhance the power of bits or qubits. We also provide a characterisation of generalised dense coding protocols, a natural subset of entanglement-assisted quantum communication protocols, finding that they can be understood as standard quantum communication protocols in real-valued Hilbert space. Though such dense coding protocols can convey up to two bits of information, we provide evidence, perhaps counter-intuitively, that resources with a small information capacity, such as a bare qutrits, can sometimes produce stronger correlations.
Along the way we leave several conjectures and conclude with a list of interesting open problems.
\end{abstract}

\maketitle

\section{Introduction}

Consider two distant parties, Alice and Bob, who each receive an input $x$ and $y$ and each produce a classical output $a$ and $b$, respectively. The joint conditional probabilities $p(a,b|x,y)$ completely characterise the correlations between Alice's and Bob's outcomes. In quantum physics, there are two basic ways through which Alice and Bob can become correlated. Either, they share prior entanglement or they directly communicate to each other. The case where Alice and Bob share prior entanglement but do not communicate corresponds to the celebrated Bell scenario \cite{Bell}. The case where they do not share entanglement but communicate corresponds, in its simplest version where communication only goes one-way, say, from Alice to Bob, to the prepare-and-measure (PM) scenario \cite{Wiesner1983}. The correlations that can be established in these two fundamental scenarios have been extensively studied qualitatively and quantitatively  \cite{Brunner_2014, Ambainis1999, Gallego2010}.

However, it is natural to also consider the more general scenario in which and Bob \emph{both} share prior entanglement \emph{and} communicate to each other, see Box~\ref{box:general-scenario}. This combination generates new types of quantum correlations, allowing for specific effects such as dense coding \cite{Bennett1992} and quantum teleportation \cite{Teleportation}. Yet, in contrast to the simpler Bell and PM scenarios, the correlations that can be established in this more general scenario have received much less attention. When communication is quantum, some studies have been made on protocols similar to quantum dense coding \cite{Abbott2018, Moreno2021}. When communication is classical, notable previous works have focused on reducing the communication complexity of specific functions \cite{doi:10.1137/S0097539797324886, Brukner2004, Zukowski2017, m2021mutually} and on Bell scenarios featuring superluminal communication \cite{Van_Himbeeck_2019, Chaves2018}.

\begin{figure}[t!]
	\centering
	\includegraphics{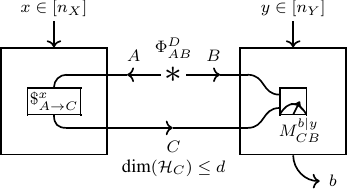}
	\caption{\textbf{Entanglement-assisted prepare-and-measure scenario:} Alice and Bob share an entangled state $\Phi^D_{AB}$, Alice gets an input $x$, applies a quantum channel $\$^{x}_{A\rightarrow C}$, and sends the resulting system $C$ to Bob. Depending on his classical input $y$, Bob performs a joint measurement $\{M^{b|y}_{CB}\}_b$ on Alice's message and his share of the entanglement and produces an outcome $b$.\label{fig:PMent}}
\end{figure}

\begin{framefloat}[detach title,before upper={\tcbtitle\quad}]{Entanglement-assisted communication scenario}
  \label{box:general-scenario}
  \begin{center}
    \includegraphics{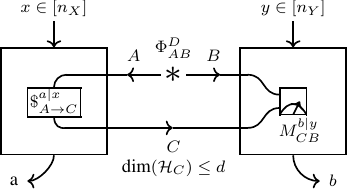}
  \end{center}
  Alice and Bob share an entangled state $\Phi_{AB}^D$ and Alice communicates a $d$-dimensional quantum system $C$ to Bob. Upon receiving her input $x$, Alice applies a quantum instrument $\{\$_{A \rightarrow \text{C}}^{a|x}\}_a$ to her part of the shared state $\Phi_{AB}^D$. She obtains a classical outcome $a$ and a $d$-dimensional system $C$ which she relays to Bob. Bob applies a joint measurement $\{M^{b|y}_{CB}\}_b$ on his part of the shared state $\Phi_{AB}^D$ and Alice's message $C$. The resulting correlations are
\begin{equation*} \label{eq:GeneralCorr}
  p(a,b \vert x,y) = \Tr \qty[ \$_{A \rightarrow \text{C}}^{a|x}[\Phi_{AB}^D]M^{b|y}_{CB}].
\end{equation*}
One can assume the communication from Alice to Bob to be classical by specialising to quantum instruments that output a system $C$ that is diagonal in a given basis. One can also consider generalisations with multi-round communication between Alice and Bob.

The case $d=1$ (no communication) corresponds to the Bell scenario. The case where $\abs{a}=1$ and $\Phi_{AB}^D$ is separable (no entanglement) corresponds to the prepare-and-measure scenario. In this work, we are interested in the \emph{entanglement-assisted prepare-and-measure scenario} where $\abs{a}=1$ and $\Phi_{AB}^D$ is arbitrary. 
\end{framefloat}

In an earlier publication \cite{companionpaper}, we considered a subclass of the general scenario of Box~\ref{box:general-scenario} in which Alice has only one possible outcome, $\abs{a}=1$, (or put more simply, she has no outcome), see Fig.~\ref{fig:PMent}. This corresponds to the \emph{entanglement-assisted (EA) prepare-and-measure scenario}: the simplest non-trivial scenario featuring both prior entanglement and quantum communication. As in a standard PM scenario, Alice receives an input $x$, prepares a quantum system $C$ depending on $x$ and sends it to Bob. Bob then performs a measurement on $C$, depending on an input $y$ and gets an outcome $b$. In a EA PM scenario, Alice and Bob additionally share a prior entangled state $\Phi_{AB}$ that can be exploited during Alice's preparation and Bob's measurement. The correlations that Alice and Bob generate can thus be written as
\begin{equation}
  \label{eq:BornRule}
  p(b |x, y) = \Tr \qty(\$^{x}_{A\rightarrow C}[\Phi_{AB}] M^{b|y}_{CB}),
\end{equation}
Where $\$^{x}_{A\rightarrow C}: L(\mathcal{H}_A) \rightarrow L(\mathcal{H}_C)$ are CPTP maps (depending on the input $x$) from Alice's share of the entangled state $\Phi_{AB}$ to a $d$-dimensional system $C$ that she communicates to Bob and $\{M^{b|y}_{BC}\}_b$ are measurements (depending on the input $y$) that Bob applies on his part of the shared state $\Phi_{AB}$ and on the system $C$ received from Alice.

The strength of the correlations that can be established between Alice and Bob depends both on the dimension $d$ of the systems communicated from Alice to Bob and on the local dimension $D$ of the shared entangled state $\Phi_{AB}$ (where possibly $D=\infty$). We refer in the following to such a resource as an $\EA_D$ qudit.

The case $D=1$ corresponds to the standard PM scenario where no entanglement is exploited and we use in this case the terminology `bare qudit'. The correlations in standard PM scenarios for different values of the dimension $d$ of the exchanged message have been extensively studied, covering a wide range of topics, including the foundations of quantum theory \cite{InfoCausality,Brassard2006}, dimension witnessing \cite{wehner2008,Gallego2010,Brunner2013}, random access coding \cite{Ambainis1999,ambainis2006,Tavakoli2015_RAC}, quantum key distribution \cite{Pawlowski2011,woodhead2015secrecy}, random number generation \cite{Li2011,li2012}, self-testing \cite{Tavakoli2018,Farkas_2019,tavakoli2020self} and certification of quantum devices \cite{Mironowicz2019,Tavakoli2020}. This has also motivated a considerable number of experiments (see e.g. \cite{Trojek_2005,Ahrens_2012,Hendrych_2012,PhysRevX.4.021047,D_Ambrosio_2014,Smania_2016}).  As shown in \cite{companionpaper}, these results and conclusions have to be revised when entanglement is allowed, i.e., when $D>1$.

\section{Summary of the results}
The present paper can be seen as a companion paper to \cite{companionpaper} aiming to explore EA PM correlations. Whereas Ref.~\cite{companionpaper} focuses on analysing and applying quantum correlations based on unrestricted entanglement, i.e.~$D=\infty$, the present paper focuses on the complementary case of fix-dimensional entanglement, i.e.~when $D$ has a specific, finite, value. The interest in such fix-dimensional scenarios is two-fold. Firstly, it is arguably natural to consider that the entanglement dimension, $D$, is known when the dimension of the channel, $d$, is known. This is certainly the case in many physical setups. Secondly, clearly bounding the entanglement restricts the possible correlations, which therefore have to be investigated separately. Indeed, many results in \cite{companionpaper} only hold when entanglement is unrestricted and conversely, as we will see, the main results obtained in this work only hold for scenarios with restricted entanglement. Let us first introduce some handy notations for the correlations produced by the different quantum and classical resources in EA PM scenarios.

\subsection{Notation}

We denote the sets of correlations obtained from quantum (classical) $d$-dimensional communication assisted by \emph{arbitrary} $D$-dimensional entanglement by $Q_d^D$ ($C_d^D$) and use the notation  $Q_d^{D^*}$ ($C_d^{D^*}$) when we specifically assume the state to be the $D$-dimensional maximally entangled state $\ket{\phi}=\frac{1}{\sqrt{D}}\sum_{i=0}^{D-1}\ket{ii}$.  When no entanglement is shared between Alice and Bob (formally equivalent to taking $D=1$), we write for simplicity $Q_d$ ($C_d$). When shared randomness is assumed to be a free resource, we consider the convex hull of these sets, which we denote $\bar{Q}_d$ ($\bar{C}_d$). When the set of correlations that can be established using resource $B$ is (strictly) contained in the set of correlations achievable with resource $A$, i.e., when resource $A$ is (strictly) more powerful than resource $B$, we write $B\subseteq A$ ($B\subset A$). When there exists given correlations that can be established with resource $A$, but not with resource $B$, we write $B<A$. Note that it is possible to have both $A<B$ and $B<A$, when in a given scenario some correlations can be established with resource $A$ but not $B$, while other correlations can be established with resource $B$ but not $A$.

\subsection{Some previously known results}
Before presenting the results specifically obtained here, let us list some basic relations between correlation sets in the EA PM scenario.

\begin{itemize}
	\item Quantum dense coding \cite{Bennett1992}, in its $d$-dimensional generalisation, allows one to double the classical capacity of a quantum channel. Hence any correlations that can be established by sending classical systems of dimension $d^2$ can be reproduced by a  $d$-dimensional quantum systems assisted by a $d$-dimensional maximally entangled state: 
	\begin{equation} \label{eq:DC}
	C_{d^2} \subseteq Q_d^{d^\ast}.
	\end{equation}

	\item The quantum teleportation protocol \cite{Teleportation} allows one to effectively send a $d$-dimensional quantum system by means of sharing a $d$-dimensional maximally entangled state and communicating a $d^2$-dimensional classical message. This implies that
	\begin{equation} \label{eq:teleportation}
	Q_d \subseteq C^{d^\ast}_{d^2}.
	\end{equation}

        \item Entanglement can enhance the correlations obtainable from one bit of classical communication \cite{doi:10.1137/S0097539797324886}, i.e.~
          \begin{equation}
            C_2 \subset C_2^2 .
          \end{equation}

	\item For binary-outcome measurements, the correlations obtained from qubit communication can be simulated with one bit of classical communication and 1 ebit \cite{hyperbits}. This implies
	\begin{equation}
	Q_2 \subseteq C_2^{2^\ast} \hspace{0.5cm} \text{(binary outcomes)}
	\label{eq:qubitsvsEAbitsBinary} .
	\end{equation}	

	\item It was recently shown \cite{frenkel2021entanglement} that 
	\begin{equation}
	C_2^{2^\ast} \subseteq \bar{C}_4.
	\end{equation}
		
	\item In the companion paper \cite{companionpaper} it was shown that
	\begin{equation}\label{oldres}
	C_3^{D}<Q_3
	\end{equation}
	for every choice of $D$.

	\item In the companion paper \cite{companionpaper} it was shown that
	\begin{equation}\label{oldres2}
	Q_{2}^2\subset Q_{2}^4 .
	\end{equation}

\end{itemize}

\begin{figure}[h!]
	\centering
	\subfloat{
		\includegraphics{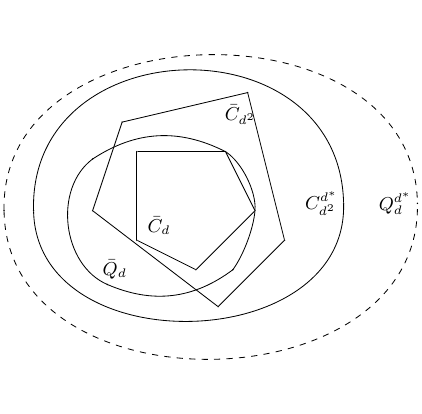}
	}
	\caption{Elementary relations from quantum teleportation \eqref{eq:teleportation} and dense coding \eqref{eq:DC} between sets of correlations for different resources. The set $Q_d^{d^\ast}$ is dotted to indicate that we do not know its relation to the set $C_{d^2}^{d^\ast}$. \label{fig:elementaryrelations}}
\end{figure}

Some of the above relations are summarised and represented in Figure~\ref{fig:elementaryrelations}.

\subsection{Results of the present paper}


\emph{(i)}. Dense coding allows one to boost the capacity of a qubit by sharing qubit entanglement ($Q_2 \subset Q_2^2$). However, the use of higher-dimensional entanglement can create even stronger correlations, in spite of the capacity of the communication not being further increased. It was shown in \cite{companionpaper} that an $\EA_4$ qubit is more powerful than an $\EA_2$ qubit, i.e., that $Q_2^2 \subset Q_2^4$. Similarly, as mentioned above, it is well known that qubit entanglement improves correlations obtained from communicating a bit, $C_2 \subset C_2^2$. In Section~\ref{SecGallegoWitness}, we show that that there are correlations that can be established by an $\EA_4$ classical bit that cannot be established when using an $\EA_2$ bit, i.e., that
	\begin{equation}
	C_2^2\subset C_2^4.
	\end{equation}


\emph{(ii)} Since the use of an $\EA_{2}$ qubit effectively leads Bob to measure a four-dimensional quantum system, any correlation produced by it can also be obtained by a bare ququart, i.e., $Q_2^2\subseteq Q_4$. In Section~\ref{SecGram} we prove that this relation is strict, i.e., that 
\begin{equation}
  \label{res}
  Q_2^{2} \subset Q_4 .
\end{equation}
A key insight in reaching this result is obtained from considering a natural subset of $\EA_{2}$ qubit strategies, which we name \emph{generalised dense coding protocols}. In these protocols the shared state is maximally entangled and Alice is restricted to performing unitary operations on her shared qubit before she sends it to Bob. We prove that the set of correlations obtained in generalised dense coding is identical to that obtained from a bare ququart restricted to a \emph{real} Hilbert space. This mostly reduces showing \eqref{res} to establishing that there are correlations in ququart communication scenarios that require a complex-valued Hilbert space.

\emph{(iii)} From dense coding,  we know that an $\EA_2$ qubit is stronger than a bare qubit, and from \eqref{res}, that it is weaker than a bare ququart, i.e., $Q_2 \subset Q_2^{2} \subset Q_4$. How does an $\EA_2$ qubit compare to a bare qutrit, which represents an intermediary resource? Since an $\EA_2$ qubit has a higher information capacity than a qutrit, it is reasonable to expected it to be a stronger resource. However, in Section~\ref{SecSICMUB}, based on numerical evidence in a specific communication task, we conjecture that this is not always be true. In other words, we provide evidence that $Q_2^2<Q_3$.

\emph{(iv)}. The relation \eqref{oldres2} was established in \cite{companionpaper} through the introduction of the flagged-RAC, a modified version of the standard Random Access Code game \cite{Ambainis1999}. In Section~\ref{SecFlaggedRAC}, we analyse the f-RAC in further detail and compute its success probability for different types of resources, both with and without entanglement.

\begin{figure}[h!]
	\centering
	\subfloat[$\EA_4$ qubits are more powerful than $\EA_2$ qubits and cannot be simulated using ququarts (section~\ref{SecFlaggedRAC}).]{
		\includegraphics{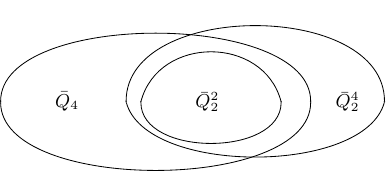}
	} \hspace{0.5cm}
	\subfloat[\emph{Conjecture}: qutrit correlations cannot be simulated using $\EA_2$ qubits (section~\ref{SecSICMUB}). Dense coding establishes the reverse; there is no strict ordering between these sets of correlations.]{
		\includegraphics{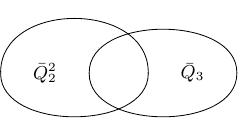}
	}
	\subfloat[Real ququarts and generalised dense coding are equivalent resources (section~\ref{sec:GDCrealququarts}). Ququarts are, however, strictly more powerful than generalised dense coding (section~\ref{SecMUBSIC}).]{
		\includegraphics{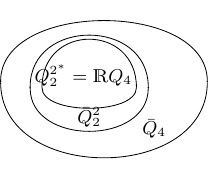}
	}
	\caption{Summary of the most notable relations between sets of correlations for different (entanglement-assisted resources) derived in this work. \label{fig:summary}}
\end{figure}

Some of the results that we obtain here are  represented in Figure~\ref{fig:summary}. They  represent only a first attempt towards systematically understanding correlations in EA PM scenarios. In Section~\ref{SecDiscussion}, we list some of the many problems that remain open.

\section{Entanglement-assisted bits: higher dimensional entanglement provides an advantage}
\label{SecGallegoWitness}

In this section, we give an initial illustration that allowing higher-dimensional entanglement can make the EA communication setting more powerful by comparing the values of a dimension witness, by Gallego \etal{} \cite{Gallego2010}, that can be attained using classical or quantum communication assisted by entanglement of different local dimensions. The witness in question is applicable to a scenario in which Alice has $n_X=5$ inputs and Bob has $n_Y=4$ inputs that yield binary outcomes, and reads
\begin{equation}
  W_{5} = \sum_{xy} c_{xy} E_{xy} ,
\end{equation}
where the correlators $E_{xy}$ are defined in terms of the probabilities by
\begin{equation}
  E_{xy} = p(b=0 | x, y) - p(b=1 | x, y)
\end{equation}
and the coefficients $c_{xy}$ are specified by the coefficient matrix
\begin{equation}
  c = \begin{pmatrix}
    1 &   1  &   1   &  1 \\
    1  &   1 &    1 &   -1 \\
    1  &   1   & -1  &   0 \\
    1  &  -1  &   0  &   0 \\
    -1   &  0  &   0 &   0
  \end{pmatrix} .
\end{equation}

In \cite{companionpaper} we derived upper bounds on this witness in the EA setting with $d = 2, 3, 4$ dimensional classical and quantum communication allowing for arbitrary entanglement. We also established that the introduction of entanglement makes a definite difference by maximising the witness using the see-saw method \cite{companionpaper} with entangled states of local dimension $D=4$ and observing that we attained values of $W_{5}$, equal to or just under the upper bounds we derived, that significantly exceeded corresponding upper bounds on $W_{5}$ that hold when there is no entanglement\footnote{These were derived analytically in \cite{Gallego2010} in the classical cases and we derived them numerically using the Navascu\'es-V\'ertesi method in the quantum case \cite{Navascues2015}.}.

To see the difference that changing $D$ makes here, we also compute here the values of $W_{5}$ that we can attain, again using the see-saw method, but this time with the entanglement limited to $D = 2$. 

To establish optimality of our lower bounds for $\EA_D$ dit communication, we look for upper bounds. Definition 2' in \cite{companionpaper} suggests a simple way this can be done using the Navascu\'es-V\'ertesi method: correlations $p(b|x,y)$ can be reproduced by an $\EA_D$ $d$-dimensional classical communication protocol if there exist $d$ unnormalised $D$-dimensional states $\{\rho^{x,c}\}_{c\in [d]}$ for each input $x$ such that $\sum_{c=1}^d \rho^{x,c} = \rho$ is independent of $x$ and $\tr(\rho)=1$ and $\abs{b}$-outcome POVM measurements $\{M_{b|y,c}\}_b$ for each input $y$ and $c \in [d]$ such that 
\begin{equation}
  p(b|x,y) = \sum_{c=1}^d \tr \bqty\Big{\rho^{c,x} M_{b|y,c}} .
\end{equation} 
By sampling such states $\rho^{c,x}$ and measurements $M_{b|y,c}$ to build the moment matrix, we can obtain bounds on $\EA_D$ dit correlations.

In general, the number of operators quickly becomes very large, making the problem computationally hard, but in the special case of one classical bit ($d = 2$), we are able to establish the optimality of our $\EA_2$ bit bound.  

\begin{table}[t!]
  \centering
  \begin{tabular}{cCCCCCC}
    \toprule
    \multicolumn{1}{c}{}& \text{C}2 & \text{Q}2 & \text{C}3 & \text{Q}3 & \text{C}4 & \text{Q}4 \\
    \midrule no ent. & 8 & 9 & 10 & 11.527 & 12 & 13.036 \\
    \cmidrule(lr){1-7} $D=2$ & 9 & 13.036 & \textcolor{blue}{11.272} & 14 & \textcolor{blue}{12.769} & 14 \\
    \cmidrule(lr){1-7} $D=4$ & 9.0343 & \textcolor{blue}{13.036} & \textcolor{blue}{11.515} & 14 & \textcolor{blue}{13.036} & 14 \\
    \cmidrule(lr){1-7} $D=\infty$ & 9.0343 & \textcolor{OliveGreen}{13.095} & \textcolor{OliveGreen}{11.563} & 14 & \textcolor{OliveGreen}{13.095} & 14 \\    \bottomrule
  \end{tabular}
  \caption{Values for the dimension witness $W_{5}$ of \cite{Gallego2010} for 2, 3, and 4 dimensional classical (C2, C3, C4) and quantum (Q2, Q3, Q4) communication with no entanglement and assisted by entanglement of local dimension $D = 2$ and $D = 4$ and of unrestricted dimension ($D = \infty$). Values in black are optimal, values in \textcolor{blue}{blue} are (possibly suboptimal) lower bounds, and values in \textcolor{OliveGreen}{green} are (possibly suboptimal) upper bounds. \label{table:Gallego}}
\end{table}

The different witness values, including the new ones we derived for $D = 2$, are summarised in Table~\ref{table:Gallego}. We see that in both the classical and quantum case, the presence of entanglement amplifies correlations substantially beyond the scenario without entanglement. In particular, we see that the task can be optimally performed ($W=14$, which is algebraically maximal) already with three-dimensional quantum communication assisted by 1 ebit, even if three-dimensional entanglement would be required to implement a dense coding protocol.

Our results furthermore indicate that higher dimensional entanglement generally improves classical communication. In the case of one classical bit, for which we know the optimal value of $W_{5}$ for $D = 2$, our results show, analogously to the quantum case that
\begin{quote}
\textit{an $\EA_{4}$ bit is more powerful than an $\EA_{2}$ bit.}	\label{eq:setineqCphi}
\end{quote}
We conjecture that similar relations hold for entanglement assisted $d$-dimensional classical (dit) communication. 

For $\EA_{2}$ qubit communication, we find that the witness value $W_{5} \approx 13.036$ can be attained using a generalised dense coding strategy, equal to the bare ququart bound up to numerical precision. Since every $\EA_{2}$ qubit state is effectively a ququart state, this bound is also optimal. In the next section we identify a correspondence that explains the coincidence of these bounds.

Finally, we note that the traditional quantum dense coding protocol is substantially suboptimal as it only achieves $W=12$ \cite{stockholm}.

\section{Entanglement-assisted qubits vs ququart communication}
\label{SecGram}

Here, we compare in more detail the relationship between $\EA_{2}$ qubits and the non-EA prepare-and-measure setting in which the source can prepare states of dimension up to four. According to Table~\ref{table:Gallego} in the previous section, we can obtain up to the same value, $W_{5} \approx 13.036$, of the Gallego witness using either $\EA_{2}$ qubits or bare ququarts. In one direction this is not surprising: because Alice can use an $\EA_{2}$ qubit to effectively prepare a ququart for Bob, clearly anything that can be done with $\EA_{2}$ qubits can be achieved using bare ququarts. The main result we establish in this section is that this inclusion is strict: ququarts are actually strictly more powerful than $\EA_{2}$ qubits. We investigate this in the following based on a connection that we noticed between a generalised version of the dense coding protocol and bare real-valued ququarts. We then present numerical evidence for a robust gap between $\EA_{2}$ qubits and ququarts, which allows one in principle to discriminate between them experimentally.

\subsection{Generalised dense coding is equivalent to real ququart communication} \label{sec:GDCrealququarts}

In quantum dense coding, Alice and Bob initially share two qubits in a maximally entangled state
\begin{equation}
  \ket{\phi_{+}} = \frac{1}{\sqrt{2}} \pqty\big{\ket{00} + \ket{11}} ,
\end{equation}
and Alice applies one of four possible unitaries
\begin{IEEEeqnarray}{c+c+c+t+c}
  U_{0} = \id, & U_{1} = \sigma_{Z} , &
  U_{2} = \sigma_{X} , &or&  U_{3} = \sigma_{Z} \sigma_{X}
  \IEEEeqnarraynumspace
\end{IEEEeqnarray}
to her qubit which she then sends to Bob. This effectively prepares one of the four orthogonal Bell states for Bob, which he can perfectly distinguish by performing a Bell state measurement. The task allows Alice to communicate two classical bits to Bob while only physically sending him one qubit. It follows that the strategy can be used to simulate any probability distribution $p(b|x,y)$ where Alice has up to $n_{X} = 4$ inputs.

More generally, we can consider a variant in which Alice may perform any qubit unitaries $U_x$ and Bob may perform any measurements but they still start with a Bell pair $\ket{\phi_{+}}$ as the initial state. We will refer to such protocols as \emph{generalised dense coding}. Interestingly, generalised dense coding turns out to be in one-to-one correspondence to a prepare-and-measure protocol, without entanglement, in which Alice communicates real ququart states to Bob. To see this, we recall that any set of pure states $\{\ket{\psi_x}\}_x$ is completely determined (up to a global rotation) by their Gram matrix $G_{xx'} \equiv \braket{\psi_x}{\psi_{x'}}$. A general qubit unitary can be written as
\begin{equation}
 U_x = \mathbf{z}_{x} \cdot \vect{\sigma} \, e^{i \varphi_x} ,
\end{equation} 
where $\varphi \in [0,2\pi)$, $\vect{\sigma} \equiv (i\mathds{1},\sigma_X,\sigma_Y,\sigma_Z)$, and $\mathbf{z}_x \in \mathds{R}^4$ is a unit vector $\norm{\mathbf{z}_x}=1$. With this parameterisation, the states
\begin{equation}
  \ket{\psi_{x}} = U_{x} \otimes \id \ket{\phi_{+}}
\end{equation}
that can be prepared by generalised dense coding have a Gram matrix of the form
\begin{IEEEeqnarray}{rCl}
  G_{xx'} &=& \bra{\phi_+} U_x^\dagger U_{x'} \otimes \mathds{1} \ket{\phi_+}
  \nonumber \\
  &=& \frac{1}{2} \Tr \bqty\big{U_x^\dagger U_{x'}} \nonumber \\
  &=& \mathbf{z}_x \cdot \mathbf{z}_{x'} \,
  e^{i(\varphi_{x'} - \varphi_{x})} , \label{eq:GramEAqubits}
\end{IEEEeqnarray}
where we used $\Tr[\sigma_i \sigma_j] = 2 \delta_{ij}$. The Gram matrix can thus always be made real by global phase transformations $\ket{\psi_x} \mapsto \ket{\psi_x} e^{i \delta_x}$ of the states, which do not change the correlations $p(b|x,y)$ that can be generated with them. Conversely, any $n_{X} \times n_{X}$ real symmetric matrix $G$ of rank at most four and with $G_{xx} = 1$ for all $x$ necessarily admits a factorisation $G_{xx'} = \mathbf{z}_{x} \cdot \mathbf{z}_{x'}$ in terms of unit vectors $\mathbf{z}_{x} \in \mathds{R}^{4}$, i.e., can be expressed in the form \eqref{eq:GramEAqubits}, and thus corresponds to a set of states $\ket{\psi_{x}}$ that could be prepared by generalised dense coding up to an overall unitary that could be performed afterwards by Bob. It follows that the set of correlations possible with generalised dense coding is identical to the set correlations possible using real ququart communication, i.e.,
\begin{quote}
  \begin{center}
    \textit{Real ququarts and generalised dense coding are equivalent resources}.
  \end{center}
\end{quote}
Since real ququart space is strictly larger than the state space of two classical bits, this  already suggests that generalised dense coding ought to give rise to stronger correlations than standard dense coding. In section~\ref{SecGallegoWitness} we have shown that this intuition is correct (see Table~\ref{table:Gallego}).

The correspondence between generalised dense coding and real ququart strategies suggests a new way to bound generalised dense coding correlations from the exterior: we can use the Navascu\'es-V\'ertesi semidefinite relaxation method \cite{Navascues2015} to bound correlations in the regular PM scenario with $d=4$, sampling only real operators. Since the hierarchy is known to converge over real Hilbert spaces \cite{PhysRevA.92.042117}, this method allows us to find tight upper bounds on generalised dense coding correlations.

\subsection{Ququarts are strictly stronger than $\EA_{2}$ qubits}
\label{SecMUBSIC}

The equivalence we have just established means that we can show that generalised dense coding and ququart communication are inequivalent resources by showing that complex ququarts are necessary to generate some correlations. Generalised dense coding, as we defined it above, is not quite as general as an $\EA_{2}$ qubit, as in the latter we explicitly allow any initial two-qubit pure state and Alice to perform any CPTP maps. However, as we will see, for the specific constructions we use this additional generality is not helpful and the inequivalence we will find between ququarts and generalised dense coding also extends to $\EA_{2}$ qubits.

To first show the inequivalence between generalised dense coding and ququarts, then, a natural approach is to identify a witness $W$ that self-tests a set of states which do not fit into the real subspace of $\mathbb{C}^{4}$, i.e., are not all real and cannot all be made real by changing the basis. Fortunately, two such self tests are already known, which we recapitulate here.

The first self-tests  a tomographically complete set of states in $L(\mathbb{C}^4)$. Ref.~\cite{Brunner2013} introduced the following prepare-and-measure scenario. Alice has $n_X=d^2$ possible inputs and Bob has binary outcomes and $n_Y=\binom{d^2}{2}$ possible inputs labelled by pairs of integers $y,y'\in[d^2]$ such that $y<y'$. The correlation witness reads 
\begin{align}
  W &\equiv \sum_{x<x'}p \bigl(b=0 \big|x,(x,x') \bigr)
      - p \bigl(b=0 \big| x',(x,x') \bigr) \nonumber \\ 
& \leq \frac{1}{2}\sqrt{d^5(d-1)^2(d+1)} \label{eq:SICW},
\end{align}
where the right-hand-side is an upper bound respected by all quantum models without entanglement in which Alice communicates a $d$-dimensional system. It was proven in \cite{Tavakoli2019} that $W$ can saturate the right-hand-side if and only if Alice's preparations form a set of symmetric informationally complete projectors (SIC). SICs (if they exist in a given dimension) are tomographically complete. For the case of our interest, namely $d=4$, existence is known and therefore the maximal value $W=48\sqrt{5}$ is attainable with complex ququarts, but not with real ququarts, and therefore not using generalised dense coding.

The second approach is to self-test mutually unbiased bases (MUBs), which play an important role in many areas of quantum information, in particular for quantum key distribution \cite{Planat_2006} and quantum state discrimination \cite{WOOTTERS1989363}. It is a well known result that there are five MUBs in $\mathds{C}^4$ (e.g., \cite{DURT_2010}), while in the restriction to the real Hilbert space $\mathds{R}^4$ only three MUBs exist \cite{boykin2005real}. We can thus distinguish generalised dense coding from ququarts if we can show that more than three MUBs are needed to obtain certain correlations in dimension four.

In \cite{PhysRevLett.121.050501,Farkas_2019} a communication task is introduced, which is essentially an $\eta^d \rightarrow 1$ QRAC \cite{Tavakoli2015_RAC} with an additionally promise. On each round, in addition to the input $y$, Bob receives an input $z \in S^{\eta}_2$, where $S^{\eta}_2$ are the possible subsets of $[\eta]$ containing two elements. The promise is that $x \in z$, i.e., on each round Bob will only be questioned about two of Alice's inputs. Bob will thus have to distinguish between two of Alice's states, but Alice does not know which two states before she sends the message. Bob must therefore be ready to distinguish any pair of states on every round of the protocol.
	
The maximum success probability for this communication task is uniquely achieved if and only if $\eta$ MUBs exist in dimension $d$. The intuition is that if $\eta$ mutually unbiased bases exist, Bob can measure in a different mutually-unbiased basis for every input $\eta$. The game then essentially reduces to a $2^d \rightarrow 1$ QRAC. Obtaining the highest success probability for this modified QRAC for $\eta=4$, $d=4$, requires the existence of four MUBs, and hence it can not be obtained with real states since only three MUBs exist in $\mathds{R}^4$. As only four MUBs are needed, we also see that tomographically complete objects are not strictly needed for an advantage compared with real ququarts or generalised dense coding.

The ability to self-test both SIC states and complex MUBs is also sufficient to show that ququarts are strictly more powerful than $\EA_{2}$ qubits, as if these tasks were achievable with $\EA_{2}$ qubits they would necessarily have to be achievable with generalised dense coding. To see this, suppose that Alice manages to prepare a set $\{\ket{\psi_{x}}\}$ of SIC or MUB states for Bob using an $\EA_{2}$ qubit. We express the states prepared for Bob by Alice as
\begin{equation}
  \label{eq:ea-prepared-mub-state}
  \ketbra{\psi_{x}} = \sum_{j} \ketbra*{\psi_{j|x}}
\end{equation}
for
\begin{equation}
  \label{eq:mub-kraus-state}
  \ket*{\psi_{j|x}} = K_{j|x} \otimes \id_{B} \ket{\psi}
  = \sum_{k} c_{k} \pqty\big{K_{j|x} \ket{k}} \ket{k} ,
\end{equation}
where the $K_{j|x}$s are Kraus operators acting $\mathcal{H}_{A} \to \mathcal{H}_{C}$ associated to the CPTP maps performed by Alice and
\begin{equation}
  \label{eq:mub-initial-state}
  \ket{\psi} = \sum_{k} c_{k} \ket{k} \ket{k}
\end{equation}
is the initial two-qubit state expressed in its Schmidt decomposition. As $\ket{\psi_{x}}$ is pure, we can infer form \eqref{eq:ea-prepared-mub-state} that $\ket*{\psi_{j|x}} \propto \ket{\psi_{x}}$ and then from \eqref{eq:mub-kraus-state} that the Kraus operators $K_{j|x}$ for all $j$ are proportional to each other on the support of Alice's marginal of $\ket{\psi}$ and thus that Alice's CPTP maps must act unitarily. Furthermore, as we assume that Alice is preparing SIC or MUB states, we can infer that the initial state must be maximally entangled. This was already shown for MUBs in \cite{Mozes_2005}. More generally, in both cases, the states $\ketbra{\psi_{x}}$ sum to an operator proportional to the identity and we thus have
\begin{equation}
  \sum_{x} \Tr_{C}\bqty\big{\ketbra{\psi_{x}}} \propto \id_{B} ,
\end{equation}
while \eqref{eq:ea-prepared-mub-state} starting with the initial state \eqref{eq:mub-initial-state} gives
\begin{equation}
  \sum_{x} \Tr_{C}\bqty\big{\ketbra{\psi_{x}}}
  \propto {c_{0}}^{2} \ketbra{0} + {c_{1}}^{2} \ketbra{1} ,
\end{equation}
which is only the same if $c_{0} = c_{1} = 1/\sqrt{2}$, i.e., if the state is maximally entangled.

If Alice manages to prepare SIC states or any number of MUBs for Bob using an $\EA_{2}$ qubit, we can thus infer that she must be accomplishing this using generalised dense coding which, as we pointed out above, is not possible for SICs or for more than three MUBs, both of which can be self tested. We conclude that
\begin{quote}
  \begin{center}
    \textit{Ququarts are a strictly stronger resource than $\EA_{2}$ qubits.}
  \end{center}
\end{quote}

\subsection{ Robustness}
\label{secSICrobustness}

Having shown that $\EA_{2}$ qubits and ququarts are inequivalent resources, the natural next step is to investigate the  robustness of this result. We do this using the SIC witness \eqref{eq:SICW} in the following\footnote{In principle, one could as well do this for the MUB witness, but this is significantly more computationally demanding.}.

Without loss of generality, we may consider effective states $\rho_x \in L(\mathbb{R}^4)$ and furthermore restrict them to being pure, i.e., $\rho_x = \ketbra{\psi_x}{\psi_x}$. Since the outcomes are binary, we can write \cite{Tavakoli2019}
\begin{equation}
  W = \max_{\ket{\psi_x}\in L(\mathbb{R}^4)}
  \sum_{x<x'} \sqrt{1 - \abs{\braket{\psi_x}{\psi_{x'}}}^2} .
\end{equation}
While we are unable to evaluate the right-hand-side analytically, it can be reliably optimised numerically by parameterising the states using hyperspherical coordinates (the space of real ququarts is a 4-sphere). We implemented several different optimisation methods and consistently find  
\begin{equation}
  \label{eq:conj}
  W \approx 106.75 .
\end{equation}
This is to be compared to the known complex ququart value $W = 48 \sqrt{5} \approx 107.33$. Notably, we were unable to improve the value \eqref{eq:conj} by allowing for higher dimensional entanglement. It would therefore be interesting to determine the maximal value of $W$ for qubit communication and unrestricted entanglement. One possible approach would be to use the methods introduced in \cite{companionpaper}, but this would be computationally demanding.

\begin{figure}[htbp]
  \centering
  \includegraphics{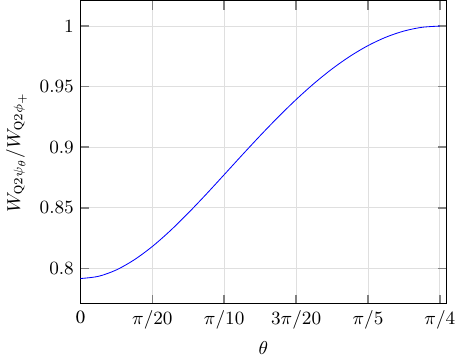}
  \caption{Heuristic bounds for the SIC witness \cref{eq:SICW} for the shared entangled state $ \ket{\psi_\theta} = \cos \theta \ket{00} + \sin \theta \ket{11}$.}
  \label{fig:SICtheta}
\end{figure}

Finally, we have also considered the optimal value of the witness \eqref{eq:SICW} when $\phi_+$ is substituted for a partially entangled pure two-qubit state of the form 
\begin{equation}  \ket{\Phi} = \ket{\psi_\theta} \equiv \cos \theta \ket{00} + \sin \theta \ket{11}, \label{eq:thetastate} \end{equation} assuming general CPTP maps.
One would expect that correlations monotonically depend on the degree of entanglement. Alternating convex searches (displayed in Figure~\ref{fig:SICtheta}) clearly indicate the expected relation. Notably, for $\theta=0$ we have a product state and the witness value reduces to that obtained with qubit communication without entanglement. This constitutes strong numerical evidence that swapping the shared maximally entangled state for partially entangled states does not affect our conclusion about the noise robustness of the SIC witness, and in fact can only increase the gap.

\section{Can qutrits outperform qubits assisted by qubit-entanglement?}
\label{SecSICMUB}

Trivially, anything that can be done with a bare qubit can also be done with an $\EA_{2}$ qubit. Is the same true when the bare qubit is replaced with a bare qutrit? Clearly, anything that can be done with a bare qutrit can also be done with a bare ququart. But we have already shown that $\EA_{2}$ qubits are strictly weaker than bare ququarts. Thus, it is seemingly nontrivial to ask whether there exists a task in which bare qutrits can outperform $\EA_2$ qubits.
	
Because an $\EA_2$ qubit is a quantum resource with a higher information capacity than a bare qutrit, one may reasonably expect the former to be more powerful. Contrary to this intuition, we here provide evidence that there exist tasks in which a qutrit can outperform an $\EA_2$ qubit.

Our candidate communication scenario is immediately inspired by the Bell inequality developed in \cite{huang2020nonlocality} and the geometric relations highlighted in that work. For this communication scenario, Alice has $n_X=9$ possible inputs and Bob has $n_Y=4$ possible inputs and ternary outputs. The correlation witness can be written in the form 
\begin{equation}
  W = \sum_{xyb} c_{xyb} \, p(b | x,y) ,  
\end{equation}
where all coefficients $c_{xyb}$ are either $+1$ or $-1$. To define them, following \cite{huang2020nonlocality}, we write Alice's input as $x=x_0x_1\in\{0,1,2\}^2$ and Bob's input as $y=y_0y_1\in\{0,1\}^2$:
\begin{equation}
  c_{xyb} = \begin{cases}
    -1 & \text{if } b = x_{y_{1}} - y_0(-1)^{y_1}x_{\bar{y}_1} \mod 3 \\
    +1 & \text{otherwise}
  \end{cases} ,
\end{equation}
where the bar-sign denotes bit-flip. Interestingly, in contrast to most quantum information games (see e.g.~Refs~\cite{PhysRevX.4.021047, Blackjack} for exceptions) this communication scenario admits an elegant interpretation in terms of the real-life colloquial game of Battleship, as detailed in Ref.~\cite{Emeriau2020} (where the game is also independently introduced).

The algebraically maximal witness value is $W=36$, which corresponds to earning one point for each pair $(x,y)$. Using the construction of \cite{huang2020nonlocality}, this value can be saturated with tomographically complete qutrit states and measurements. Let Alice's nine state form the Hesse SIC which (up to normalisation) is
\begin{align}\label{eq:sic}\nonumber
  & \ket{\psi_{00}}\sim \begin{pmatrix}
  0\\1 \\-1
  \end{pmatrix} & \ket{\psi_{01}}\sim \begin{pmatrix}
  0\\1 \\-\omega
  \end{pmatrix}  && \ket{\psi_{02}}\sim\begin{pmatrix}
  0\\1 \\-\omega^2
  \end{pmatrix} \\\nonumber
  & \ket{\psi_{10}}\sim \begin{pmatrix}
  -1\\0 \\1
  \end{pmatrix}  & \ket{\psi_{11}}\sim \begin{pmatrix}
  -\omega\\0 \\1
  \end{pmatrix}  && \ket{\psi_{12}}\sim \begin{pmatrix}
  -\omega^2\\0 \\1
  \end{pmatrix}\\
  & \ket{\psi_{20}}\sim \begin{pmatrix}
  1\\-1 \\0
  \end{pmatrix} & \ket{\psi_{21}}\sim \begin{pmatrix}
  1\\-\omega \\0
  \end{pmatrix} && \ket{\psi_{22}}\sim \begin{pmatrix}
  1\\-\omega^2 \\0
  \end{pmatrix},
\end{align}
where $\omega=e^{2\pi i/3}$. Let Bob perform measurements in four mutually unbiased bases, given (up to normalisation) by
\begin{align}\label{eq:mubs}
  &\begin{bmatrix}
    1 & 0 & 0\\
    0 & 1 & 0\\
    0 & 0 & 1
  \end{bmatrix} ,
  &\begin{bmatrix}
    1 & 1 & 1 \\
    1 & \omega & \omega^2 \\
    1 & \omega^2 & \omega
  \end{bmatrix} ,
  &&\begin{bmatrix}
    1 & \omega & \omega\\
    \omega & 1 & \omega\\
    \omega & \omega & 1
  \end{bmatrix} ,
  &&\begin{bmatrix}
    1 & \omega^2 & \omega^2\\
    \omega^2 & 1 & \omega^2\\
    \omega^2 & \omega^2 & 1
  \end{bmatrix} .
\end{align}
Here, the columns represent the basis states and each matrix represents measurement $y=1,2,3,4$ (equivalently, $00,01,10,11$) respectively. A simple calculation of the Born rule then shows that for every $(x,y)$, the probability of outputting the value of $b$ that is associated to $c_{xyb}=-1$ is zero. Thus, we find that we can attain the algebraically maximal value
\begin{equation}
  \label{eq:qutritvalue}
  W = 36. 
\end{equation}

The question becomes whether this optimal value can also be achieved using generalised dense coding, i.e.~using real ququarts.  We used alternating convex searches to optimise $W$ for real ququarts. The largest witness value we find is $W \approx 35.42$, which is clearly less than \cref{eq:qutritvalue}. In addition, we find that the witness value we attain decreases monotonically with the degree of entanglement for general qubit entanglement (again assuming arbitrary CPTP maps), similar to what we saw for the SIC witness in \cref{fig:SICtheta}.

Unfortunately, it proved too computationally expensive to derive an upper bound on $W$ using the Navascu\'es-V\'ertesi method \cite{Navascues2015}, so it remains an open question whether the numerical gap reported here truly cannot be closed.

Finally, we mention that by qubit communication assisted by four-dimensional entanglement, we found by numerical search that it is possible to recover the qutrit value \eqref{eq:qutritvalue}.

\section{The f-RAC game: bounds with and without entanglement}
\label{SecFlaggedRAC}

In this section, we generalise the communication game originally introduced in \cite{companionpaper} to prove that qubit communication assisted by higher-dimensional entanglement can outperform protocols with only qubit-entanglement to a family of communication tasks and discuss optimal strategies for different communication resources. We now introduce a simple communication task that reveals this advantage using the minimal number of preparations. Specifically, we have $n_X =5$ inputs for Alice and $n_Y=3$ measurement settings for Bob with binary outcomes.

The starting point for our task is the usual $2\rightarrow 1$ quantum random access code (QRAC) \cite{Ambainis} in which Alice has $n_X=4$ preparations which she encodes into a qubit that is sent to Bob. Bob has $n_Y=2$ possible inputs with binary outcomes. With his first setting he aims to distinguish between $x\in \{1,2\}$ and $x\in\{3,4\}$ while with his second setting he aims to distinguish between the inputs $x\in\{1,3\}$ and $x\in\{2,4\}$. When Alice and Bob share a maximally entangled qubit state, they can perfectly perform this task using a dense coding protocol.

In our modification of this task, we introduce a fifth input $x=5$ for Alice and a third input $y=3$ for Bob. The fifth state serves as a flag that is relevant only for the third measurement. With his third input, Bob attempts to determine whether he received the flag or any of the bit-encoding states used in the RAC (i.e., distinguish $x\in\{1,2,3,4\}$ and $x=5$). To quantify how well this task can be performed, we {consider a linear function
	\begin{IEEEeqnarray}{rCl}
	\label{eq:frac}
	W &=& \pqty{E_{11} + E_{21} - E_{31} - E_{41}} \nonumber \\
	&&+\> \pqty{E_{12} - E_{22} + E_{32} - E_{42}} \nonumber \\
	&&+\> \beta \pqty{E_{13} + E_{23} + E_{33} + E_{43}} - \gamma E_{53} ,
	\end{IEEEeqnarray}
	with $E_{xy} = p(b=1 | x,y) - p(b=-1 | x,y)$,
	where we have introduced two tuneable parameters $\beta,\gamma \geq 0$ which determine the relative weight given to the identification of the flag. Notice that for $\gamma=0$, the third setting of Bob is trivialised and the task reduces to a standard QRAC.
	
	\begin{figure}[t!]
		\centering
		\includegraphics{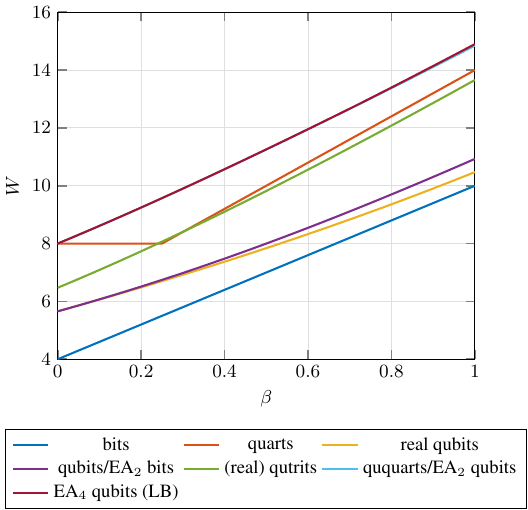}
		\caption{ Optimal value/lower bounds (LB) of {the witness \eqref{eq:frac}} for different $\beta$ {and $\gamma = 4\beta$} using different resources, with and without shared entanglement. Note that the ququart bound (which is equal to the bound from generalised dense coding), is almost overlapping with the lower bound we find for qubits assisted by four-dimensional entanglement. The latter is nevertheless higher, and the gap increases {with $\beta$}; see \cref{fig:Wqqq2e} for a more detailed picture. \label{fig:Wdifferentresources}}
	\end{figure}
	
	\begin{figure}[t!]
		\centering
		\includegraphics{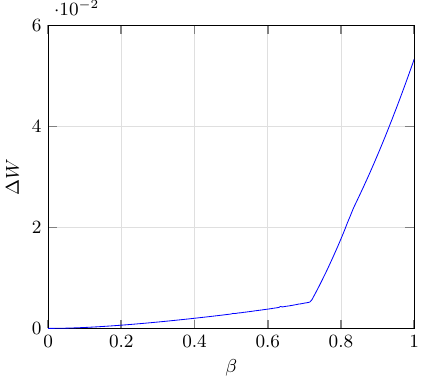}
		\caption{The difference between the optimal values of $W$ for ququarts and $\EA_2$ qubits as a function of $\beta$. \label{fig:Wqqq2e}}
	\end{figure}
	
	We focus on the parameter regime for which the discrimination associated to $y=3$ has equal weight, i.e.~$\gamma=4\beta$, and consider different resources for the task. The results are summarised in Figure~\ref{fig:Wdifferentresources}, and discussed separately below.

	\subsection{Classical communication}
	
	{In the bare setting, the classical bound on the f-RAC witness \eqref{eq:frac} (with $\gamma = 4\beta$) can be derived by maximising it over all classical deterministic communication strategies with messages limited to a given dimension $d$. We consider $d=2$ (bits) and $d=4$ (two bits) explicitly here. Since the number of such strategies is manageably small, we can do this by checking all of them on a computer, while leaving $\beta$ a free variable.}
	
	{When Alice is only allowed to send one bit ($d=2$), the maximal value of the witness is
		\begin{equation}
		W = 4 + 6 \beta .
		\end{equation}
		An} optimal strategy that attains this is to use the communicated bit to distinguish between $x\in\{1,2,3\}$ and $x\in\{4,5\}$. When Bob receives $0$, he outputs $+1$ for every value of $y$ while when he receives $1$ he outputs $-1$ for every value of $y$.

	When Alice is allowed to send 2 classical bits ($d=4$), the maximal value of the witness is
	\begin{equation}
	W = \begin{cases}
	8 &\text{if}~\beta \leq 1/4\\
	6 + 8\beta &\text{if}~\beta \geq 1/4
	\end{cases}.
	\end{equation}

	
	The following strategy (which depends on the value of $\beta$) attains this
	\begin{itemize}
		\item If $\beta \leq 1/4$, Alice uses the following encoding:
		\begin{IEEEeqnarray}{r"c"l} \label{eq:enc}
		1 \rightarrow 00 \,, & 2 \rightarrow 01 \,, & 3 \rightarrow 10, \nonumber\\ 4 \rightarrow 11 \,, & 5 \rightarrow 11 \,. 
		\end{IEEEeqnarray} Upon receiving Alice's message Bob uses the following decoding function: 
		\begin{alignat}{2}  00 &\rightarrow (+1, +1, +1), \,\,\,\,\,\,\,\,\,\,\,   && 01 \rightarrow (-1, +1, +1),  \nonumber \\
		10 &\rightarrow (+1, -1, +1),  && 11 \rightarrow (-1, -1, +1),
		\end{alignat} where for every possible message from Alice, we wrote Bob's outputs for different $y$ as an ordered tuple.
		\item If $\beta \geq 1/4$, Alice again uses the  encoding \cref{eq:enc} but upon receiving Alice's message Bob outputs the following 
		\begin{alignat}{2}  
		&00 \rightarrow (+1, +1, +1), \,\,\,\,\,\,\,\,\,\,\, && 01 \rightarrow (-1, +1, +1), \nonumber \\ &10 \rightarrow (+1, -1, +1), && 11 \rightarrow (\cdot, \cdot, -1), \end{alignat}
	\end{itemize}
	where the dots in the last line indicate that the response-function of Bob in this case is irrelevant because it has no weight in the witness.
	
	Thus, in contrast to the qubit case, there is a qualitative change in the optimal strategy at a critical value of $\beta$. We will find a similar situation when comparing qubits to four-dimensional quantum systems.

	\subsection{Quantum communication}
	
	When the communication from Alice to Bob is quantum, we can write
	\begin{equation}
	E_{xy} = \Tr[\rho_{x} M_{y}],
	\end{equation}
	where $M_y \equiv M_{1|y}-M_{-1|y}$ and $\rho_x$ are quantum states of dimension $d$.
	
	For quantum communication we considered the case of $d=2,3,4$. We used the semidefinite relaxation hierarchy of  \cite{Navascues2015} to find upper bounds on the witness which were confirmed to be tight by matching them to lower bounds obtained by alternating SDPs.
	
	We find that convergence requires a high relaxation level, which we can achieve by implementing the symmetrisation techniques of \cite{Tavakoli2019}. Our tight upper bounds are obtained with the relaxation level $\rho + E $  $+E^2 + \rho^2 +\rho E + E \rho +\rho^3 +E^3+\rho E \rho + EE \rho + E \rho E $ for the considered cases of $d=2,3,4$. They equal the lower bounds up to solver precision.
	
	For the symmetrisation, note that the function $W$ is invariant under the Dihedral group $\mathcal{D}_4$. This non-abelian permutation group of order eight has four generators: $\mathds{1}$, $g_1 = \{(12)(34), E_{x2} \rightarrow -E_{x2}\}$, $g_2 = \{(13)(24), E_{x1} \rightarrow -E_{x1}\}$, $g_3= \{(23), E_{x1} \leftrightarrow E_{x2}\}$, where we used cycle notation to denote permutations of the states. From these generators we construct the full group: $\mathcal{D}_4 = \{\mathds{1},g_1,g_2,g_3,g_1g_2=g_2g_1,g_1g_3 = g_3g_2,g_2g_3 = g_3g_1, g_1g_2g_3\}$, where symmetries act from left to right.

	\subsubsection{Qubits}
	
	For qubits we attempt to guess an explicitly strategy and prove this strategy is optimal by showing that it saturates the upper bounds obtained using the hierarchy methods \cite{Navascues2015}. 
	
	The strategy builds on intuition from Quantum Random Access Codes (QRAC), which correspond to the special case of $\beta = 0$, for which the optimal qubit encoding is to take four states that form a square on a disk of the Bloch sphere e.g. $\ket{0}, \ket{1}, (\ket{0} \pm \ket{1})/\sqrt{2}$. Starting from this intuition, we can extend the strategy to our witness where the relevance of the flag grows with $\beta$. We may always fix the flag state such that its Bloch vector is along the positive $z$-axis ($\rho_5=\ketbra{0}{0}$). Choose the square configuration for the four preparations $x=1,2,3,4$ in the $xy$-plane (perpendicular to the flag). Now, the more we decrease the overlap between the square configuration and the flag, the better we may distinguish the flag but the worse we perform at the QRAC. The value of $\beta$ determines the optimal tradeoff between these two effects (see an illustration in Figure~\ref{fig:blochvecs}).
	
	\begin{figure}[t!]
		\centering
		\subfloat[Optimal qubit states for $\beta = 0.3$]{
			\includegraphics{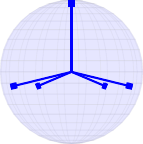}
		} \hspace{1.5cm}
		\subfloat[Optimal qubit states for $\beta = 1$]{
			\includegraphics{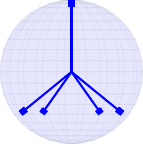}
		}
		\caption{Bloch sphere for optimal qubit states for different values of $\beta$. The north-pole state is the flag and the remaining four states form a square in a (non-unit) disk of the Bloch sphere. For $\beta=0$, the square configuration is placed in the $xy$-plane. As $\beta$ increases, the square configuration moves towards the south-pole in an increasingly small disk of the Bloch sphere.} \label{fig:blochvecs}
	\end{figure}
	
	Inspired by this geometrical picture and symmetry considerations, we may take 
	\begin{alignat}{4}
	\mathbf{r}_1 &= (\,\,\,0 &&,  \sin \theta &&, -\cos \theta &&)^T, \\
	\mathbf{r}_2 &= (\sin \theta &&, \,\,\,0&&, -\cos \theta &&)^T, \\
	\mathbf{r}_3 &= (-\sin \theta &&, \,\,\,0&&, -\cos \theta &&)^T, \\
	\mathbf{r}_4 &= (\,\,\,0 &&, -\sin \theta &&,-\cos \theta &&)^T, \\
	\mathbf{r}_5 &= (\,\,\, 0 &&,\,\,\,0&&, \,\,\,1 &&)^T,
	\end{alignat}
	where $\mathbf{r}_x$ are Bloch vectors, i.e. we write $\rho_x = \frac{\mathds{1} + \mathbf{r}_x\cdot \bm{\sigma}}{2}$. By plugging these ansatze into $W$ and using the best value of $\theta$ we find
	\begin{IEEEeqnarray}{rCl}
        \label{eq:SgamequbitW}
	W &=& \norm{\mathbf{r}_1 + \mathbf{r}_2 - \mathbf{r}_3 - \mathbf{r}_4}
	+ \norm{\mathbf{r}_1 - \mathbf{r}_2 + \mathbf{r}_3 - \mathbf{r}_4}
	\nonumber \\
	&&+\> \beta \norm{\mathbf{r}_1 + \mathbf{r}_2
          + \mathbf{r}_3 + \mathbf{r}_4 - 4 \mathbf{r}_5} \nonumber \\
        &=& 4\sqrt{2} \sin \theta + 4\beta (1 + \cos \theta)  \nonumber \\
        &=& 4 \beta + 4 \sqrt{2 + \beta^{2}},
	\end{IEEEeqnarray}
	where in the first line we assumed that the optimal measurements are non-degenerate and in the last line we use that $\max_{\theta} \pqty\big{A \cos(\theta) + B \sin(\theta)} = \sqrt{A^{2} + B^{2}}$, which is saturated when
	\begin{equation}
	\beta \tan \theta = \sqrt{2}.
	\end{equation}
	
	For $\beta = 0$, we recover the optimal QRAC strategy, which achieves $W = 4\sqrt{2}$. For $\beta > 0$ the lower bound \cref{eq:SgamequbitW} matches up to numerical precision the upper bound we find using the hierarchy methods of \cite{Navascues2015}.

	\subsubsection{Ququarts}
	
	For ququarts, we look for the optimal strategy for different ranges of $\beta$ using alternating convex searches to match our upper bounds obtained using the methods of \cite{Navascues2015}.
	In contrast to the qubit case, the optimal four-dimensional quantum strategy does not seem to admit a simple geometric interpretation.  For relatively small $\beta$ and for large $\beta$, however, the optimal states and measurements take a simple form. For $0 \leq \beta \lesssim 0.835$, we start by fixing $\ket{\psi_5}= \frac{1}{2}(\ket{00} + \ket{01} + \ket{10} + \ket{11})$ for the flag state when performing numerical searches. The third measurement is then given by
	\begin{equation}
	M_{3}^{*} = \frac{1}{2} \pqty\big{
		\mathds{1} \otimes \mathds{1} - \sigma_X \otimes \sigma_X
		- \sigma_X \otimes  \mathds{1} - \mathds{1} \otimes \sigma_X } .  
	\end{equation}
	We find that
	\begin{align}
	M_1^\ast &= \sigma_Z \otimes \mathds{1}, \\
	M_2^\ast &= \mathds{1} \otimes \sigma_Z,
	\end{align}
	are optimal. Having fixed the measurements, we can straightforwardly compute the optimal remaining states as an eigenvalue problem. This leads to
	\begin{equation}
	W = 4\norm{M_1^\ast+M_2^\ast +\beta M_3^\ast}_\infty + 4\beta.
	\end{equation}
	Computing the eigenvalues of $M_1^\ast+M_2^\ast +\beta M_3^\ast$, using Mathematica find that  the final result can be put in the form 
\begin{equation} W = 4\beta + \frac{4}{\sqrt{3f(\beta)}}\sqrt{(8+3\beta^2)f(\beta)+2f(\beta)^2+12\beta^2+8 } \end{equation}
where
\begin{equation} f(\beta) = {\left(9\,\beta^2+3\,\sqrt{3}\,\beta\,\sqrt{-8\,\beta^4-13\,\beta^2-16}-8\right)}^{1/3} \, . \end{equation}
	At $ \beta \approx 0.835 $, there a phase transition, but the strategy stays close to optimal until $\beta \gtrsim 1$ (at worst for $\beta =1$ it is $10^{-3}$ lower than our tight upper bound), at which point we again observe a phase transition.
	
	For $\beta \gg 1$, it is clear that there will be a regime where the contribution from the third measurement dominates the contributions coming from the other two measurements corresponding to the QRAC. In this regime, it becomes essential for Bob to distinguish the fifth state (the flag) from the other states so it becomes advantageous to take the first four states $\rho_1,\dots,\rho_4$ in a subspace orthogonal to the flag state, $\rho_5$. By doing this, the contribution to the witness from the third measurement is maximised, $\beta \norm{\rho_1 +  \rho_2 + \rho_3 +  \rho_4 - 4\rho_5}_1 = 8 \beta$. For ququarts this means that the first three states are in a three dimensional subspace. At this point the optimal strategy takes a particularly simple form because the problem reduces to finding the optimal strategy for a the QRAC in a three-dimensional subspace in order to maximise the contribution of the first two measurements. The optimal success probability for such a task was proven using semidefinite relaxations in \cite{Navascues2015} and analytically in \cite{Tavakoli2018}. In terms of our observable, this implies
	\begin{IEEEeqnarray}{rCl}
	\IEEEeqnarraymulticol{3}{l}{
		\norm{\rho_1 + \rho_2 - \rho_3 - \rho_4}_{1}
		+ \norm{\rho_1 - \rho_2 + \rho_3 - \rho_4}_{1}} \nonumber \\
	\qquad &=& 8(2p_{\text{max}} -1) \nonumber \\
	&=& 2(1+ \sqrt{5}),
	\end{IEEEeqnarray}
	where $p_\text{max}$ is the maximal success probability for the QRAC.
	
	Choosing the flag state $\ket{\psi_5} =  \ket{3}$, the following strategy saturates this bound 
	\begin{align}
	\ket{\psi_1} &= \ket{0}, \label{eq:qqa} \\
	\ket{\psi_2} &= \frac{1}{\sqrt{2}}\qty(\ket{1}+\ket{2}), \\
	\ket{\psi_3} &= \frac{1}{\sqrt{2}}\qty(\ket{1}-\ket{2}), \\
	\ket{\psi_4} &= \ket{1} \label{eq:qqb}.
	\end{align}
	Taking everything together, the optimal witness value for $\beta \gg 1$ becomes 
	\begin{equation}
	W = 2(1+\sqrt{5})+8\beta.
	\end{equation}
	This is equal up to numerical precision to our lower bound for $\beta \gtrsim  2(1+\frac{1}{\sqrt{5}})$. 
	
	\subsection{Entanglement-assisted communication}
	
	\subsubsection{Entanglement-assisted classical communication.}

	Using the methods of \cite{Navascues2015}, we could prove that our heuristic bounds $\EA_{2}$ bits are optimal and equal to the qubit bounds. Moreover, in contrast to the witness in section~\ref{SecGallegoWitness}, we do not find an improvement using higher dimensional entanglement using numerical searches.

	\subsubsection{Entanglement-assisted quantum communication.}
	
	We found real ququart strategies that reach the (complex) ququart bound. Given the correspondence between real ququart communication and generalised dense coding derived in section~\ref{SecGram}, this means that there exist generalised dense coding strategies that can simulate the ququart correlations. Because all $\EA_{2}$ qubit states can be effectively prepared by sending bare ququarts, these strategies are also optimal. Importantly, however, using alternating convex search, we find that for all values of $\beta$ there exist $\EA_{4}$-qubit strategies that outperform generalised dense coding protocols as well as ququart communication. 
	
	In other words,
	
	\begin{quote}
		\begin{center} \textit{$\EA_4$ qubits are a strictly more powerful  than $\EA_2$  qubits,} \end{center}  
	\end{quote}
	and
	\begin{quote}
		\begin{center}
			\textit{$\EA_4$ qubits are a strictly more powerful than ququarts for some tasks}.  \end{center}
	\end{quote}
	
	In Figure~\ref{fig:Wqqq2e} we illustrate how the advantage of higher-dimensional entanglement over quantum dense coding  increases with $\beta$ (corresponding to the regime where identification of the fifth state (the flag) becomes important). 
	
	We present strategies involving four-dimensional shared entanglement that outperform generalised dense coding for all $\beta$. In the regime where this advantage becomes largest, $\beta \gg 1$, it was proven in \cite{companionpaper} to be optimal. For all $\beta$, we find strategies where Alice and Bob initially share two copies of the maximally entangled two-qubit state,
        \begin{equation}
          \ket{\phi}_{AB} = \ket{\phi_+}_{A_1B_1} \otimes \ket{\phi_+}_{A_2B_2} .
        \end{equation}
        Consequently, the states held by Bob after Alice's communication are of dimension eight, corresponding to a qubit system (the communication) and the ququart system (Bob's share of $\ket{\phi}_{AB}$). 
	
	Alice's strategy consists in applying a two-qubit unitary $U_x$ on $A_1A_2$, discarding the first qubit $A_1$ and sending the second qubit $A_2$ to Bob. The channel she implements is thus given by 
	\begin{equation}
	\includegraphics{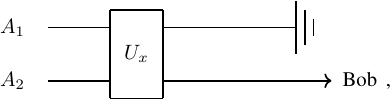}
	\end{equation}
	
	It can be checked that the effective states $\rho_x$ Bob measures in his laboratory are then rank-2 states of the form
	\begin{equation}
	\rho_x = \frac{1}{2} \pqty\big{
		\ketbra{\psi_x}{\psi_x} + \ketbra{\varphi_x}{\varphi_x}} .
	\end{equation}
	
	In general, it is difficult to find a closed-form expression for the optimal unitaries $U_x$ for Alice. In the large $\beta$ regime, however, the problem is significantly simplified. As was the case without entanglement, for $\beta\gg 1$ it becomes advantageous to take the fifth state orthogonal to the first four states. Based on the intuition for the $d=4$ strategy, we look for states for which
	\begin{equation}
	\trnorm{\rho_5 - \rho_i} = \trnorm{\rho_1 - \rho_4}
	= \trnorm{\rho_2 - \rho_3} = 2 ,
	\end{equation}
	where $i=1,\dots,4$ and which additionally maximise 
	\begin{equation}
	\trnorm{\rho_1 -\rho_2} = \trnorm{\rho_1 -\rho_3}
	= \trnorm{\rho_4-\rho_2} = \trnorm{\rho_4-\rho_3} .
	\end{equation}
	The following choice of unitaries works,
	\begin{align}
	U_1 &= \openone \otimes \openone,
	\label{eq:u1} \\
	U_2 &= CNOT_1 \, CNOT_2, \\
	U_3 &= \openone \otimes \sigma_X \, CNOT_1 \, CNOT_2, \\
	U_4 &= \openone \otimes \sigma_Z, \label{eq:u4} \\
	U_5 &= \openone \otimes \sigma_Z\sigma_X \, CNOT_2,
	\end{align}
	where $CNOT_i$ is the controlled-$NOT$ gate with the control on the $i$th qubit. With this choice, we find
        \begin{equation}
          W = 2(2+ \sqrt{2}) + 8\beta.
        \end{equation}  Our lower bounds also suggest our large $\beta$ strategies for both $d=4$ in eqs.~\eqref{eq:qqa}--\eqref{eq:qqb} and the entanglement assisted qubits described here become optimal already for relatively small $\beta$, at which point the gap becomes large, see \cref{fig:Wbeta0to4}.  
	
	\begin{figure}[t!]
		\centering
		\includegraphics{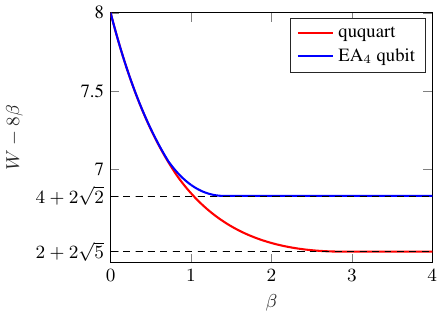}
		\caption{Best bounds found by numerical search for $0 \leq \beta \leq 4$.  \label{fig:Wbeta0to4}}
	\end{figure}

\section{Outlook and open problems}
\label{SecDiscussion}

Prepare-and-measure scenarios have been extensively studied in the past decade, but the usual analysis assumes that the communicating parties may only be classically correlated. On the other hand, Bell scenarios famously reveal the power of entanglement to enhance correlations, but no communication is allowed between the parties. Nevertheless, dense coding and teleportation show that entanglement may radically amplify correlations for quantum and classical communication. While these protocols establish some basic relations between the different classical/quantum sets of correlations with and without entanglement, surprisingly little is known beyond that in the context of general prepare-and-measure scenarios with shared entanglement. 

In this work and in the companion paper \cite{companionpaper}, we lay the groundwork for studying this question more systematically. Here, we have presented several elementary results, but these only scratch the surface of the topic. We conclude with a list of some of the many interesting open problems.

\subsection{Open problem related immediately to the conjectures and results of this work}

\begin{enumerate}
	\item We have found that generalised dense coding based on qubits has a one-to-one correspondence with communication of real ququarts. Is it possible to map more general classes of entanglement-assisted prepare-and-measure scenarios  to standard prepare-and-measure scenarios with suitably restricted Hilbert spaces?\\
	
	\item  Is it always the case that the maximally entangled state is the best to enhance classical or quantum communication, i.e., does it hold that $Q_d^{D} \subseteq Q_d^{D^\ast}$ and $C_d^{D} \subseteq C_d^{D^\ast}$? We conjecture that the answer is negative, as we find numerical evidence in \cref{SecGallegoWitness} for $	\bar{C}_2^{4^{ \ast}} < C_2^{4}$.	\\
	
	\item We numerically evidenced a robust gap between correlations obtained from ququarts and correlations obtained from $\EA_2$ qubits (see Section~\ref{secSICrobustness}). Can one prove this gap? A positive answer would enable experiments to discriminate between the two resources. It would also be interesting to identify prepare-and-measure scenarios that allow for a larger gap between these two resources.\\
	
	\item We showed that $Q_2^2 \subset Q_4$. Is it also true that $Q_d^d \subset Q_{d^2}$ for any $d$? We conjecture yes.\\
	
	\item We showed that $C_2^2 \subset C_2^4$. Is it also true that $C_d^d \subset C_{d}^D$ for any $d$ and some $D>d$? We conjecture yes. And what about the case of quantum communication? \\
	
	\item We conjectured that correlations obtained from a qutrit can outperform correlations obtained from an $\EA_2$ qubit (see Section~\ref{SecMUBSIC}). Can one prove this result? If yes, is it also true (for some other communication game) that $Q_2^D<Q_3$ for any $D$?\\	
	
	\item  Our conjecture that qutrits can outperform an $\EA_2$ qubit implies that quantum resources with a smaller classical capacity can sometimes outperform quantum resources with a larger classical capacity. This raises the question whether there exists an entangled state that cannot be used to increase the classical capacity of the communication channel, but which can nevertheless be used to gain other communication advantages?

\end{enumerate}

\subsection{Generic open problems}

\begin{enumerate}
	\item Is there some bound on the entanglement dimension above which no advantage is obtained in generating correlations? In other words, does there exist some large $D$ such that $\forall k:Q^D_d = Q^{D+k}_d$ for all finite $D$, and similarly when the communication is classical? A more general version of this question is a long-standing open question in communication complexity \cite{Buhrman2010}.\\
	
	\item For a given communication task, is it possible to identify the minimal entanglement dimension $D$ required in order to generate the optimal correlations using quantum or classical communication? How can one bound this minimal dimension?	\\
	
	\item Does every entangled state enable an advantage over bare communication in some communication task? In other words, is it true that $Q_d < Q_d^{\psi}$ for every state $\psi$ and some $d$? Does the analogous hold for classical communication?\\
	
	\item A more restricted, but still interesting, version of the previous problem is to decide whether there exists an entangled state with positive partial transpose for which $Q_d< Q_d^{\psi}$ (alternatively $C_d< C_d^{\psi}$).\\

	\item  Generally, it is interesting to enquire about the set relations between $C_d^D$ and $Q^{D'}_{d'}$. Note that already without entanglement, there are many open questions here. For instance, while $Q_2 < \bar{C}_4$ for projective measurements, we do not know the relation between $Q_2$ and $\bar{C}_D$ for arbitrary measurements. Another example is to decide whether there is a strict relationship between an EA dit and a bare qudit. Based on the result \eqref{oldres}, it seems plausible that neither $C_d^\infty$ or $Q_d$ are contained in each other.

	\item What is the relation between correlations obtained from ququart communication and the correlations obtained from qubit communication assisted by any number of ebits (i.e.~$n$ copies of the maximally entangled state, $\phi_+^{\otimes n}$)? Is one a subset of the other?\\

	\item A number of works have explored entanglement-assisted classical communication tailored to solve tasks tailored on Bell inequalities \cite{Brukner2004, Pawlowski2010, Zukowski2017, Brukner2020}. Could higher-dimensional entanglement help to outperform strategies based on the violation of a Bell inequality? Note that results presented in \cite{companionpaper} show a positive answer for the case of ternary outcomes.
\end{enumerate}

\subsection{Broader open problems}

We conclude with two larger open questions:
\begin{enumerate}
	\item We focused on Bell-scenarios with only one outcome for Alice, $\abs{a}=1$, i.e. the entanglement-assisted prepare-and-measure scenario. It would be interesting to study the most general case, when Alice may have non-trivial outcomes.
	\item In this work, we limited the communication of Alice by imposing a bound on the dimension $d$. It would be interesting to extend this to other semi-device-independence assumptions, for example an energy bound \cite{VanHimbeeck2017}, a bound on the information content \cite{Tavakoli2020informationally, InformationCorr2} or a bound on the distrust \cite{tavakoli2021semideviceindependent}. 
\end{enumerate}

\section*{Acknowledgements}

We acknowledge Nicolas Gisin's paper ``Bell inequalities: many questions, a few answers'' \cite{ref:g2009} for inspiring the title of our work. This work was supported by the Swiss National Science Foundation (Early PostDoc Mobility fellowship P2GEP2 194800), the EU Quantum Flagship project QRANGE and the Fonds National de la Recherche Scientifique F.R.S.-FNRS (Belgium) under a FRIA grant. S.P. is a Senior Research Associate of the Fonds de la Recherche Scientifique -- FNRS\@.

\bibliography{references}

\end{document}